\documentclass[12pt,twoside,a4paper]{article}
\pdfoutput=1

\usepackage{amsmath, amsthm, amsfonts, amssymb}
\usepackage{graphicx}
\usepackage{float}
\usepackage[width=\textwidth,font=small,labelfont=bf,
labelsep=endash]{caption}
\usepackage[numbers,sort&compress]{natbib}

\normalsize
\long\def\@makecaption#1#2{%
  \vskip\abovecaptionskip
  \sbox\@tempboxa{\small #1. #2}%
  \ifdim \wd\@tempboxa >\hsize
    \small #1. #2\par
  \else
    \global \@minipagefalse
    \hb@xt@\hsize{\hfil\box\@tempboxa\hfil}%
  \fi
  \vskip\belowcaptionskip}

\newcommand\Poincare {Poincar\'e\ }

\newcommand\csch {\mathrm{csch}}

\newcommand\tr {\mathrm{tr}}

\newcommand\arccosh {\mathrm{arccosh}}

\renewcommand{\thesection}{\arabic{section}}
\renewcommand{\thesubsection}{\arabic{section}.\arabic{subsection}}
\renewcommand{\theequation}{\arabic{section}.\arabic{equation}}

\begin{document}

\title{On the Holographic Entanglement Entropy for Non-smooth Entangling Curves in AdS$_4$}
\author{Georgios Pastras$^1$}
\date{$^1$NCSR ``Demokritos'', Institute of Nuclear and Particle Physics\\15310 Aghia Paraskevi, Attiki, Greece\\\texttt{pastras@inp.demokritos.gr}}

\vskip .5cm

\maketitle

\abstract{We extend the calculations of holographic entanglement entropy in AdS$_4$ for entangling curves with singular non-smooth points that generalize cusps. Our calculations are based on minimal surfaces that correspond to elliptic solutions of the corresponding Pohlmeyer reduced system. For these minimal surfaces, the entangling curve contains singular points that are not cusps, but the joint point of two logarithmic spirals one being the rotation of the other by a given angle $\delta \varphi$. It turns out that, similarly to the case of cusps, the entanglement entropy contains a logarithmic term, which is absent when the entangling curve is smooth. The latter depends solely on the geometry of the singular points and not on the global characteristics of the entangling curve. The results suggest that a careful definition of the geometric characteristic of such a singular point that determines the logarithmic term is required, which does not always coincide with the definition of the angle. Furthermore, it is shown that the smoothness of the dependence of the logarithmic terms on this characteristic is not in general guaranteed, depending on the uniqueness of the minimal surface for the given entangling curve.}


\setcounter{footnote}{0}

\def\thefootnote{\arabic{footnote}}

\newpage
\tableofcontents

\setcounter{equation}{0}
\section{Introduction}
\label{sec:introduction}

The Ryu-Takayanagi conjecture \cite{Ryu:2006bv,Ryu:2006ef} (see also \cite{Hubeny:2007xt,Nishioka:2009un,VanRaamsdonk:2009ar,VanRaamsdonk:2010pw,Takayanagi:2012kg,Blanco:2013joa,Wong:2013gua}) is a deep link between theories interrelated through the holographic duality. The conjecture relates quantitatively quantum entanglement in the boundary CFT to the geometry of the bulk theory. Considering a closed surface $\partial A$ (the entangling surface) on the AdS boundary separating it to regions $A$ and $A^C$, the associated entanglement entropy (EE)
\begin{equation}
{S_{\rm{EE}}} :=  - \tr{\rho _A}\ln {\rho _A},\quad {\rho _A} = \tr_{{A^C}}\rho 
\end{equation}
is connected to the area of an extremal co-dimension two open surface in the bulk geometry, whose boundary coincides with the entangling surface $\partial A$. More quantitatively, the entanglement entropy associated to region $A$ is given by
\begin{equation}
{S_{\rm{EE}}} = \frac{1}{4{G_N}} \, {\rm Area} \left(A^{{\rm{extr}}} \right) \, ,
\label{eq:RT_conjecture}
\end{equation}
where ${A^{{\rm{extr}}}}$ is the extremal co-dimension two surface in the bulk. This striking conjecture has opened many paths to the understanding of the emergence of gravity in theories with holographic duals, as well as to the understanding of the role of entanglement in strongly coupled systems, following the opposite direction of the duality.

The area of the minimal surface is divergent as one could easily guess due to the divergence of the bulk metric at the boundary. Introducing a radial cut-off scale $L$, the entanglement entropy for a spherical entangling surface of radius $R$ in the boundary of $AdS_{d+1}$ is given by the expressions
\begin{equation}
{S_{\rm{EE}}} \simeq
\begin{cases}
{a_1}{\left( {R/L} \right)^{d - 2}} + {a_3}{\left( {R/L} \right)^{d - 4}} +  \ldots  + {a_{d - 2}} {R/L}  + {a_0}, & d\;\rm{odd},\\
{a_1}{\left( {R/L} \right)^{d - 2}} + {a_3}{\left( {R/L} \right)^{d - 4}} +  \ldots  + {a_{d - 3}}{\left( {R/L} \right)^2} + {a_0} {\rm ln}
 {R/L} , & d\;\rm{even}.
\end{cases}
\label{eq:eeterms}
\end{equation}
The first and most divergent term is the usual ``area law'' term \cite{Bombelli:1986rw,Srednicki:1993im}, which is proportional to the area of the entangling surface. When $d$ is even, the logarithmic term is universal and it is connected to the conformal anomaly \cite{Ryu:2006bv,Ryu:2006ef,Casini:2011kv,Myers:2010xs,Myers:2010tj,Solodukhin:2008dh}. When $d$ is odd, the constant term is universal and it obeys a holographic ``c-theorem'' \cite{Myers:2010xs,Myers:2010tj}.


The coefficients $a_i$ of the expansion \eqref{eq:eeterms} strongly depend on the geometric features of the entangling surface. It has been shown \cite{Bueno:2015xda} that the presence of non-smooth points in the entangling surface has some even more significant consequences in the form of the expansion of the entanglement entropy \eqref{eq:eeterms}, as new terms arise. These terms are not dependent on the global characteristics of the entangling surface, but rather only on the features of the non-smooth points. This kind of terms is particularly interesting, as they are universal, and they are connected to the central charge of the boundary CFT theory \cite{Bueno:2015xda,Bueno:2015rda,Bueno:2015qya,Bueno:2015lza}.

Focusing to the case of AdS$_4$, the expansion of the entanglement entropy with the cut-off radial scale $L$ for an arbitrary smooth entangling curve reads, 
\begin{equation}
{S_{\rm{EE}}} = {c_1}\left( {L_0/L} \right) + {c_0} + \mathcal{O}\left( {{\left( {L_0/L} \right) ^{ - 1}}} \right),
\end{equation}
where $L_0$ is some characteristic scale of the entangling curve. The linear term is the ``area law'', in this case a ``length law'', since the entangling surface is a one-dimensional closed curve, whereas the constant term is the universal one, which is independent of the renormalization scheme. However, when the entangling curve has non-smooth cusps, a logarithmic term emerges,
\begin{equation}
{S_{\rm{EE}}} = {c_1}\left( {L_0/L} \right)  + a\ln \left( {L_0/L} \right) + {c_0} + \mathcal{O}\left( {{\left( {L_0/L} \right) ^{ - 1}}} \right) .
\end{equation}
The coefficient $a$ of the logarithmic term depends solely on the angular openings $\Omega$ of the cusps of the entangling curve and it is universal.

An obstacle to the study of the Ryu-Takayanagi conjecture is the limited number of analytically calculable open minimal surfaces in AdS geometries. The non-linearity of the equations specifying an extremal surface limits the majority of the literature to the study of minimal surfaces corresponding to spherical entangling surfaces.
Recently, non-trivial minimal surfaces have been explicitly constructed in AdS$_4$ \cite{Pastras:2016vqu,Pastras:2017afl}. These constructions are based on the inversion of Pohlmeyer reduction for the specific class of elliptic solutions of the reduced integrable system, namely the Euclidean cosh-Gordon equation. The elliptic minimal surfaces in general correspond to entangling curves with some singular non-smooth points, thus, they are an appropriate tool for the study of the contributions of such points to the entanglement entropy and its dependence on the geometric properties of the singular points.

The structure of this paper is as follows. In section \ref{subsec:minimal_surfaces}, we review some basic features of the elliptic minimal surfaces in AdS$_4$. In section \ref{sec:logarithmic_spikes}, we calculate the holographic entanglement entropy for elliptic minimal surfaces and identify the terms that emerge due to the existence of the singular points. In section \ref{sec:properties}, we study the properties of these terms and identify the geometric feature of the singular points that determines their contribution to the entanglement entropy. In section \ref{sec:discussion}, we discuss our results. Finally, there is an appendix with the properties of the elliptic minimal surfaces that are used throughout the main text.

\setcounter{equation}{0}
\section{Review of Elliptic Minimal Surfaces in AdS$_4$}
\label{subsec:minimal_surfaces}

The elliptic minimal surfaces comprise a two-parameter family of minimal surfaces. A convenient choice for these two parameters are the quantities $E$ and $\wp \left( a_1 \right)$, as defined in \cite{Pastras:2016vqu}. The first one determines the corresponding solution of the Pohlmeyer reduced system. The second one corresponds to the choice of a surface among the members of an associate (Bonnet) family of minimal surfaces, i.e. a family of surfaces that have identical principal curvatures, and, thus, the same Pohlmeyer counterpart.

In global coordinates, the elliptic minimal surfaces accept the following parametric form in terms of isothermal coordinates $u$ and $v$
\begin{align}
r &= \Lambda \sqrt {\frac{{\wp \left( u \right) - \wp \left( {{a_1}} \right)}}{{\wp \left( {{a_2}} \right) - \wp \left( {{a_1}} \right)}}{{\cosh }^2}\left( {{\ell _1}v + {\varphi _1}\left( u \right)} \right) - 1} , \label{eq:Properties_radial_global} \\
\theta  &= {\tan ^{ - 1}}\left( {\sqrt {\frac{{\wp \left( u \right) - \wp \left( {{a_1}} \right)}}{{\wp \left( u \right) - \wp \left( {{a_2}} \right)}}} \csch\left( {{\ell _1}v + {\varphi _1}\left( u \right)} \right)} \right) ,\\
\varphi  &= {\ell _2}v - {\varphi _2}\left( u \right) .
\end{align}
All quantities in the above equations, as well as the moduli of the Weierstrass elliptic and related functions are expressed in terms of the parameters $E$ and $\wp \left( a_1 \right)$. More information is provided in the appendix and in \cite{Pastras:2016vqu}.

The minimal surface intersects the AdS boundary when $u \to 2 n \omega_1$, where $n \in \mathbb{Z}$. Therefore, an appropriately anchored to the boundary minimal surface is spanned by
\begin{equation}
u \in \left(2 n \omega_1 , 2 \left( n + 1 \right) \omega_1 \right) ,\quad
v \in \mathbb{R} ,
\label{eq:boundary_spanning}
\end{equation}
where $n \in \mathbb{Z}$.

The corresponding entangling curve, is the union of two spiral curves of the form
\begin{align}
\cot {\theta _ + } &= \sinh \left( {\omega\left( {{\varphi _ + } + {\varphi _0}} \right)} \right) ,\\
\cot {\theta _ - } &= \sinh \left( {\omega\left( {{\varphi _ - } + {\varphi _0} - \delta \varphi } \right)} \right) ,
\end{align}
where $\omega$ and $\delta \varphi$ are functions of $E$ and $\wp \left( a_1 \right)$ (See appendix \ref{sec:formulae}). In \Poincare coordinates, the two curves above assume of the form of two logarithmic spirals.

\begin{figure}[ht]
\centering
\begin{picture}(80,29)
\put(0,0){\includegraphics[width = 0.30\textwidth]{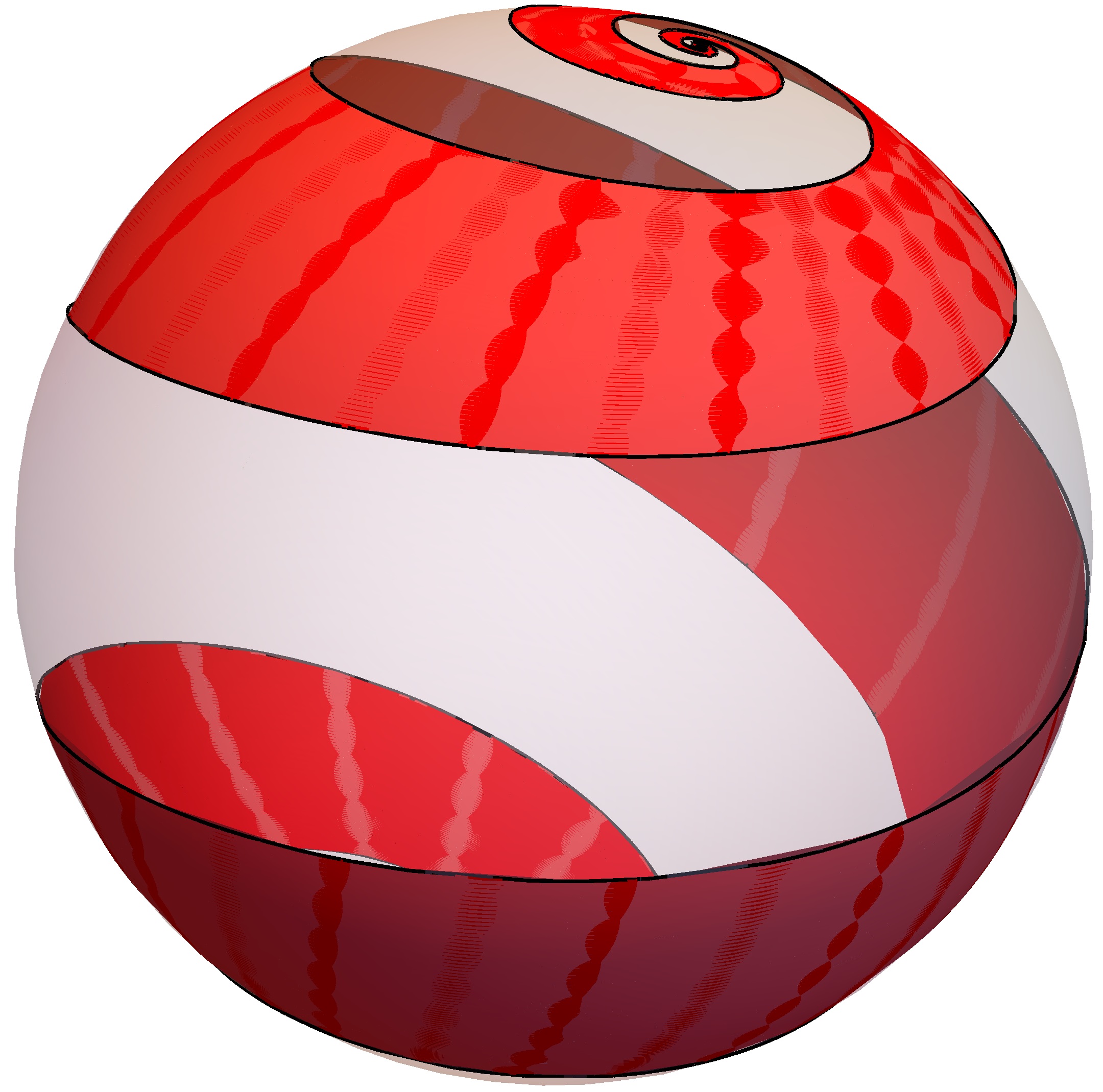}}
\put(35,0){\includegraphics[width = 0.45\textwidth]{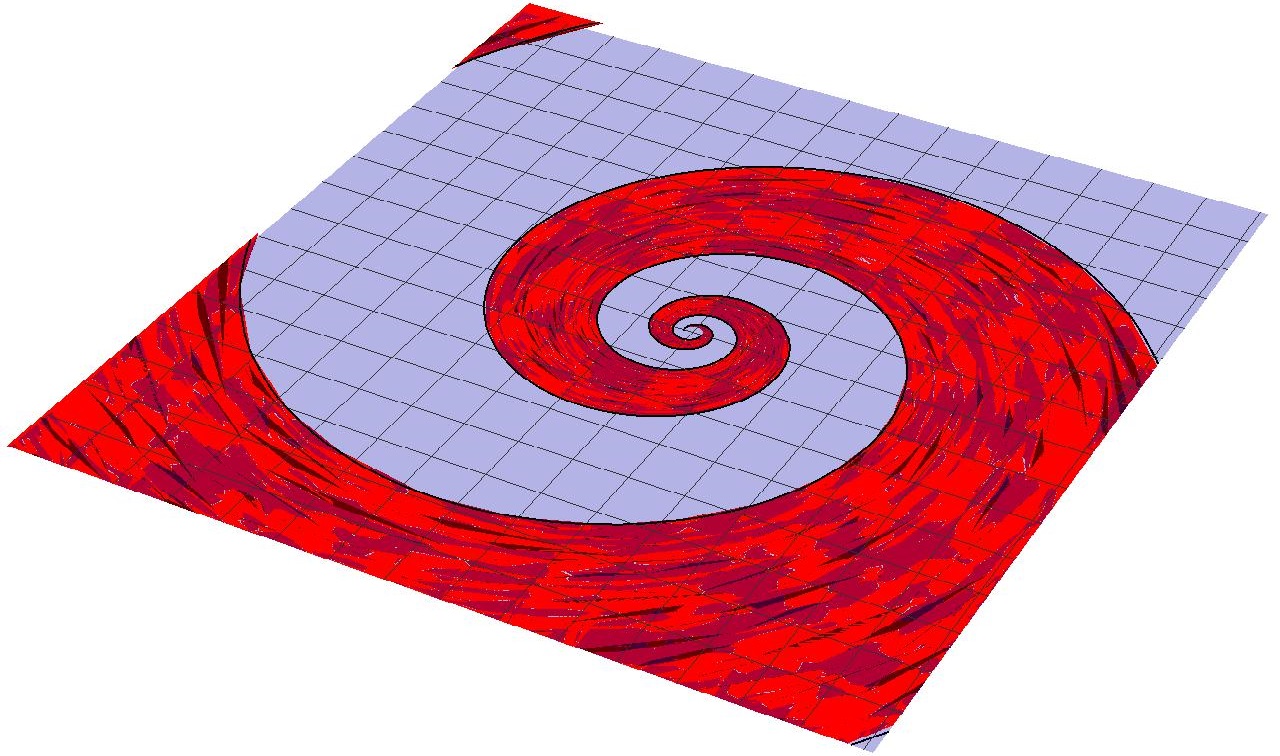}}
\end{picture}
\caption{The entangling curve and the corresponding boundary regions in global (left) and \Poincare (right) coordinates}
\label{fig:boundary_region}
\end{figure}

The parameters $\omega$ and $\delta \varphi$ completely determine the form of the entangling curve. The latter has a simple geometrical meaning. If one of the two curves comprising the entangling curve is rotated about axis $z$ by an angle $\delta \varphi$ then, it would coincide with the other. Notice that $\delta \varphi$ may be larger that $2\pi$, but then the minimal surface has self-intersections; in the following, we will not consider such cases. The form of the entangling curve and the corresponding boundary regions are depicted in figure \ref{fig:boundary_region}.

The entangling curve separates the AdS boundary to two regions of unequal size. The ratios of the area of each of these two regions to the total area of the boundary are $\delta \varphi  / 2 \pi $ and $1 - \delta \varphi / 2 \pi $. In this sense, the parameter $\delta \varphi$ plays the same role as the angle in the case of cusps. However, the two curves comprising the entangling curve meet in a non-smooth way that does not look like a cusp of finite angle $\Omega$, but to a cusp of vanishing angle, i.e. a spike. This follows from the fact that the length of a segment of the entangling curve between any given point and the singular point is infinite. Actually, the angular opening at the singular point is not well defined, since the direction of the two curves at the singular point is also not well defined. There are two exceptions to this rule. The first one is the $\omega \to 0$ limit, where the elliptic surfaces degenerate to catenoids for whom the entangling curve is disconnected but smooth. The second one is the $\omega \to \infty$ limit, where the elliptic surfaces degenerate to simple cusps with $\Omega = \delta \varphi$. In the following, we will call the non-smooth points of the general elliptic minimal surface as ``spiral spikes''.

\setcounter{equation}{0}
\section{``Spiral Spike'' Contributions to EE}
\label{sec:logarithmic_spikes}

\subsection{Terms Emerging from Non-smooth Points}

As we have already commented in section \ref{sec:introduction}, in AdS$_4$, if an entangling curve contains non-smooth cusps, the associated entanglement entropy will acquire a logarithmic term that is absent when the entangling curve is smooth \cite{Bueno:2015xda}. The coefficient of this term depends solely on the angular openings of the non-smooth points; it does not depend at all on the rest of the geometry of the entangling curve.

This logarithmic term must obey some properties that emerge from the geometry of the entangling curve. A cusp of angular opening $\Omega$ is obviously identical to a cusp of angular opening $2 \pi - \Omega$. Thus, it is expected that the coefficient $a\left( \Omega  \right)$ of the logarithmic term has the symmetry property
\begin{equation}
a\left( \Omega  \right) = a\left( {2\pi  - \Omega } \right) .
\label{eq:a_log_symmetry}
\end{equation}
Furthermore, at the limit $\Omega \to \pi$, the cusp ceases to exist, the entangling curve becomes smooth and so the logarithmic term vanishes,
\begin{equation}
a\left( \pi \right) = 0.
\label{eq:a_log_smooth_limit}
\end{equation}
The symmetry property \eqref{eq:a_log_symmetry} combined with the limit \eqref{eq:a_log_smooth_limit} implies that if the coefficient $a$ is a smooth function of $\Omega$, then
\begin{equation}
a\left( {\pi  + \Omega } \right) = \mathcal{O}\left( {{\Omega ^2}} \right) .
\label{eq:a_log_at_pi}
\end{equation}

Finally, it turns out that for small angles, the coefficient of the logarithmic term diverges as \cite{Bueno:2015xda}
\begin{equation}
\mathop {\lim }\limits_{\Omega  \to 0} a\left( \Omega  \right) \simeq \frac{\kappa }{\Omega }.
\label{eq:a_log_at_zero}
\end{equation}

The existence of the pair of spiral spikes in the entangling curves of the elliptic minimal surfaces is expected to generate non-trivial terms in the entanglement entropy that do not appear in the case of smooth entangling curves. Such terms could include a logarithmic term in a similar manner to the case of cusps. Additionally, bearing in mind that in the case of cusps the coefficient of the logarithmic term diverges for small cusp angular openings, it would not be surprising to discover a more divergent term (e.g. $L \ln L$), due to the spiky nature of the non-smooth points. Before proceeding to calculating the entanglement entropy for the elliptic minimal surfaces, we will use purely geometric arguments to deduce some basic properties of the coefficients $a$ of the terms under study, in analogy to the relations \eqref{eq:a_log_symmetry}, \eqref{eq:a_log_smooth_limit} and \eqref{eq:a_log_at_pi}.

Similarly to the case of cusps, entangling curves corresponding to $\delta \varphi$ and $2 \pi - \delta \varphi$ are identical and thus the coefficient $a$ of a term emerging due to the existence of a spiral spike must obey,
\begin{equation}
a\left( {\delta \varphi } \right) = a\left( {2\pi  - \delta \varphi } \right) ,
\end{equation}

Similarities to the case of cusps are limited to the above symmetry property. In the case of elliptic minimal surfaces, the entangling curve does not have a smooth limit as $\delta \varphi \to \pi$. At this limit, the elliptic minimal surfaces degenerate to a helicoid which is also characterized by the existence of spiral spikes. Consequently, we should not expect that $a$ vanishes at this limit in general,
\begin{equation}
a\left( \pi \right) \neq 0 .
\end{equation}
Whether the derivative of the coefficient $a$ with respect to $\delta \varphi$ at $\delta \varphi = \pi$ is vanishing, as in the case of cusps, is a more complicated question that we will face later.

Finally, there is no obvious reason to expect an expansion for small $\delta \varphi$ similar to equation \eqref{eq:a_log_at_zero}.

\subsection{Holographic EE for Elliptic Minimal Surfaces}

We may now proceed to derive the entanglement entropy for the elliptic minimal surfaces and specify the terms related to the existence of the spiral spikes. The area of the minimal surface is given by \cite{Pastras:2016vqu}
\begin{equation}
A = {\Lambda ^2}\int {dvdu\left( {\wp \left( u \right) - {e_2}} \right) } ,
\end{equation}
where $e_2$ is the intermediate root of the cubic polynomial associated with the Weierstrass elliptic function and is given by equation \eqref{eq:Elliptic_e2}. We introduce a radial cut-off $L$. Then, the cut-off area is given by
\begin{equation}
A\left( L \right) = {\Lambda ^2} \int\limits_{r < L} {dudv\left( {\wp \left( u \right) - {e_2}} \right)} .
\end{equation}

The radial coordinate $r$ is given by equation \eqref{eq:Properties_radial_global} implying that for any given value of the coordinate $u$, the inequality $r < L$ is equivalent to
\begin{equation}
v_- \left( u \right) < v < v_+ \left( u \right) ,
\end{equation}
where
\begin{equation}
{v_ \pm } \left( u \right) = \frac{1}{{{\ell _1}}}\left[ { - {\varphi _1}\left( u \right) \pm \arccosh\sqrt {\frac{{\wp \left( {{a_2}} \right) - \wp \left( {{a_1}} \right)}}{{\wp \left( u \right) - \wp \left( {{a_1}} \right)}}\left( {\frac{{{L^2}}}{{{\Lambda ^2}}} + 1} \right)} } \right] .
\end{equation}
Therefore, the cut-off area is given by
\begin{equation}
A\left( L \right) = \frac{2 {\Lambda ^2}}{{{\ell _1}}}\int_{{u_{\min }}}^{{u_{\max }}} {du\left( {\wp \left( u \right) - {e_2}} \right)\arccosh\sqrt {\frac{{\wp \left( {{a_2}} \right) - \wp \left( {{a_1}} \right)}}{{\wp \left( u \right) - \wp \left( {{a_1}} \right)}}\left( {\frac{{{L^2}}}{{{\Lambda ^2}}} + 1} \right)} } ,
\label{eq:area_1}
\end{equation}
where ${u_{\min }}$ and ${u_{\max }}$ are the values of $u$ that set the range of $v$ to zero, namely,
\begin{equation}
\wp \left( {{u_{\min / \max}}} \right) = \wp \left( {{a_2}} \right) + \left( {\wp \left( {{a_2}} \right) - \wp \left( {{a_1}} \right)} \right)\frac{{{L^2}}}{{{\Lambda ^2}}} .
\label{eq:u_min_max}
\end{equation}
As the Weierstrass elliptic function is periodic, the equation \eqref{eq:u_min_max} does not uniquely determine ${u_{\min}}$ and ${u_{\max}}$. Assuming that the range of the coordinate $u$ spanning the whole minimal surface is $\left(2 n \omega_1 , 2\left( n + 1 \right) \omega_1 \right)$, the appropriate selection for the quantities $u_{\min}$ and $u_{\max}$ is unique and their values would obey
\begin{equation}
u_{\min} \in \left(2 n \omega_1 , \left( 2 n + 1 \right) \omega_1 \right) ,\quad u_{\max} \in \left(\left( 2 n + 1 \right) \omega_1 , \left( 2 n + 2 \right) \omega_1 \right) ,
\end{equation}
so that
\begin{equation}
u_{\min} + u_{\max} = 2 \left( 2 n + 1 \right) \omega_1 .
\end{equation}

Since, the only function of $u$ appearing in the expression \eqref{eq:area_1} for the cutoff area is the Weierstrass elliptic function $\wp$, one may shift $u$ by an integer multiple of $2 \omega_1$, so that both $u_{\min}$ and $u_{\max}$ lie within $\left( 0 , 2 \omega_1 \right)$, and, thus, they obey $u_{\min} + u_{\max} = 2 \omega_1$. Furthermore, taking advantage of the fact that $\wp$ is even and periodic, we may express equation \eqref{eq:area_1} as
\begin{equation}
A\left( L \right) = \frac{4 {\Lambda ^2}}{{{\ell _1}}}\int_{{u_{\min }}}^{{\omega_1}} {du\left( {\wp \left( u \right) - {e_2}} \right)\arccosh\sqrt {\frac{{\wp \left( {{a_2}} \right) - \wp \left( {{a_1}} \right)}}{{\wp \left( u \right) - \wp \left( {{a_1}} \right)}}\left( {\frac{{{L^2}}}{{{\Lambda ^2}}} + 1} \right)} } .
\label{eq:area_2}
\end{equation}

By parts integration of the above expression yields
\begin{multline}
A\left( L \right) =  - \frac{4 {\Lambda ^2}}{{{\ell _1}}}\left( {\zeta \left( {{\omega _1}} \right) + {e_2}{\omega _1}} \right)\arccosh\sqrt {\frac{{\wp \left( {{u_{\min }}} \right) - \wp \left( {{a_1}} \right)}}{{\wp \left( {{\omega _1}} \right) - \wp \left( {{a_1}} \right)}}} \\
 - \frac{2 {\Lambda ^2}}{{{\ell _1}}}\int_{{u_{\min }}}^{{\omega _1}} {du\left( {\zeta \left( u \right) + {e_2}u} \right)\frac{{\frac{{\wp '\left( u \right)}}{{\wp \left( u \right) - \wp \left( {{a_1}} \right)}}}}{{\sqrt {\frac{{\wp \left( {{u_{\min }}} \right) - \wp \left( u \right)}}{{\wp \left( {{u_{\min }}} \right) - \wp \left( {{a_1}} \right)}}} }}} .
 \label{eq:EE_expansion_der_1}
\end{multline}
In the above formula, we may consider as the parameter of expansion the quantity ${\wp \left( {{u_{\min }}} \right)}$ instead of $L$, which is of order $L^2$. It is clear that only the first term of \eqref{eq:EE_expansion_der_1} may provide a logarithmic term; the integral can only provide polynomial terms. The expansion of the integral in powers of $L$ cannot be directly carried out, as the latter appears in both the integrand and the limits of integration. This problem can be bypassed by performing the change of variable,
\begin{equation}
x = \frac{{\wp \left( {{u_{\min }}} \right) - \wp \left( u \right)}}{{\wp \left( {{u_{\min }}} \right) - \wp \left( {{\omega _1}} \right)}} .
\end{equation}
Then, the cut-off area assumes the form
\begin{multline}
A\left( L \right) =  - \frac{4 {\Lambda ^2}}{{{\ell _1}}}\left( {\zeta \left( {{\omega _1}} \right) + {e_2}{\omega _1}} \right)\arccosh\sqrt {\frac{{\wp \left( {{u_{\min }}} \right) - \wp \left( {{a_1}} \right)}}{{\wp \left( {{\omega _1}} \right) - \wp \left( {{a_1}} \right)}}} \\
 + \frac{2 {\Lambda ^2}}{{{\ell _1}}}\int_0^1 {dx\left\lbrace {\zeta \left( {{\wp ^{ - 1}}\left( {\left( {1 - x} \right)\wp \left( {{u_{\min }}} \right) + x\wp \left( {{\omega _1}} \right)} \right)} \right) }\right.}\\
{\left. { + {e_2}{\wp ^{ - 1}}\left( {\left( {1 - x} \right)\wp \left( {{u_{\min }}} \right) + x\wp \left( {{\omega _1}} \right)} \right)} \right\rbrace\frac{{\frac{{\wp \left( {{u_{\min }}} \right) - \wp \left( {{\omega _1}} \right)}}{{\left( {1 - x} \right)\wp \left( {{u_{\min }}} \right) + x\wp \left( {{\omega _1}} \right) - \wp \left( {{a_1}} \right)}}}}{{\sqrt {x\frac{{\wp \left( {{u_{\min }}} \right) - \wp \left( {{\omega _1}} \right)}}{{\wp \left( {{u_{\min }}} \right) - \wp \left( {{a_1}} \right)}}} }}} .
\end{multline}
It is a matter of algebra to show that
\begin{multline}
A\left( L \right) = \frac{{2{\Lambda ^2}}}{{{\ell _1}}}\sqrt {\wp \left( {{u_{\min }}} \right)} \int_0^1 {\frac{{dx}}{{\sqrt {x\left( {1 - x} \right)} }}} \\
- \frac{{4{\Lambda ^2}}}{{{\ell _1}}}\left( {\zeta \left( {{\omega _1}} \right) + {e_2}{\omega _1}} \right)\ln \frac{{2 \sqrt {\wp \left( {{u_{\min }}} \right)} }}{{\sqrt {\wp \left( {{\omega _1}} \right) - \wp \left( {{a_1}} \right)} }} + \mathcal{O} \left( {\frac{1}{{\sqrt {\wp \left( {{u_{\min }}} \right)} }}} \right)
\end{multline}
or in terms of the radial cutoff $L$,
\begin{multline}
A\left( L \right) = 2\pi \Lambda \sqrt {\frac{{{\omega ^2} + 1}}{{{\omega ^2}}}} L \\
- 4{\Lambda ^2}\sqrt {\frac{{1 - {\omega ^2}}}{{3{e_2}{\omega ^2}}}} \left( {\zeta \left( {{\omega _1}} \right) + {e_2}{\omega _1}} \right)\ln \frac{2L}{{\Lambda }}\sqrt {\frac{{{\omega ^2} + 1}}{{{\omega ^2} + \left( {1 - {\omega ^2}} \right)\frac{{{e_2} - {e_3}}}{{3{e_2}}}}}}  + \mathcal{O} \left( {{L}^{-1}} \right) .
\end{multline}

The first term is the usual ``area law'' term \cite{Bombelli:1986rw,Srednicki:1993im}. The second term is the universal logarithmic term emerging from the ``spiral spikes'' of the entangling surface. Despite the spiky nature of the singular points, it turns out that no terms more divergent than the logarithmic one are present. The coefficient of the logarithmic term is
\begin{equation}
a = - 4{\Lambda ^2}\sqrt {\frac{{1 - {\omega ^2}}}{{3{e_2}{\omega ^2}}}} \left( {\zeta \left( {{\omega _1}} \right) + {e_2}{\omega _1}} \right) .
\label{eq:coef_a}
\end{equation}

In the cusp limit ($\omega \to \infty$), the entanglement entropy equals
\begin{equation}
A\left( L \right) = 2\pi \Lambda L + 4{\Lambda ^2}\sqrt { - \frac{1}{{3{e_2}}}} \left( {\zeta \left( {{\omega _1}} \right) + {e_2}{\omega _1}} \right)\ln \frac{L}{{2\Lambda }}\sqrt {\frac{{3{e_2}}}{{2{e_2} + {e_3}}}}  + O\left( {{L}^{-1}} \right) ,
\end{equation}
implying that the logarithmic term coefficient equals
\begin{equation}
a = 4{\Lambda ^2}\sqrt { - \frac{1}{{3{e_2}}}} \left( {\zeta \left( {{\omega _1}} \right) + {e_2}{\omega _1}} \right) .
\end{equation}

\setcounter{equation}{0}
\section{Properties of the Logarithmic Term}
\label{sec:properties}

\subsection{Small $\delta \varphi$ Limit}
\label{subsec:small_angle}
It is not difficult to acquire an asymptotic expression for the coefficient of the logarithmic term \eqref{eq:coef_a} in the limit of small values of the parameter $\delta \varphi$, in order to compare with equation \eqref{eq:a_log_at_zero}. The angle $\delta \varphi$ gets arbitrarily small at the limit $E \to 0$. At this limit, both $a_1$ and $a_2$ approach $\omega_3$, which is the sum of the real and purely imaginary half-periods of the associated Weierstrass elliptic function. The properties of the elliptic minimal surfaces \eqref{eq:construction_cosh_a1}, \eqref{eq:construction_cosh_a2} and \eqref{eq:boundary_omega} imply that
\begin{equation}
\wp \left( {{a_1}} \right) = {e_2} - \frac{1}{{1 - {\omega ^2}}}\frac{E}{2},\quad \wp \left( {{a_2}} \right) = {e_2} + \frac{{{\omega ^2}}}{{1 - {\omega ^2}}}\frac{E}{2} .
\end{equation}
Expanding the Weierstrass elliptic function around $\omega_3$ yields
\begin{equation}
\wp \left( {{\omega _3} + x} \right) = {e_2} - \frac{1}{{4{\Lambda ^4}}}{x^2} + \mathcal{O} \left( {{x^3}} \right) ,
\end{equation}
which in turn implies that
\begin{align}
{a_1} &= {\omega _3} - 2{\Lambda ^2}\sqrt {\frac{1}{{1 - {\omega ^2}}}\frac{E}{2}}  + \mathcal{O}\left( {{E^{3/2}}} \right) , \\
{a_2} &= {\omega _3} - 2i{\Lambda ^2}\sqrt {\frac{{{\omega ^2}}}{{1 - {\omega ^2}}}\frac{E}{2}}  + \mathcal{O}\left( {{E^{3/2}}} \right) .
\end{align}
Substituting the above in the formula for the parameter $\delta \varphi$ \eqref{eq:boundary_df}, we get
\begin{equation}
\delta \varphi  = 4{\Lambda ^2}\zeta \left( {{\omega _1};{\Lambda ^{ - 4}},0} \right)\sqrt {\frac{1}{{1 - {\omega ^2}}}\frac{E}{2}} \left( {\omega  + \frac{1}{\omega }} \right) + \mathcal{O}\left( {{E^{3/2}}} \right) .
\end{equation}

Similarly, expanding the formula providing the coefficient of the logarithmic term \eqref{eq:coef_a} yields
\begin{equation}
a = - 4{\Lambda ^2}\zeta \left( {{\omega _1};{\Lambda ^{ - 4}},0} \right)\sqrt {\frac{{1 - {\omega ^2}}}{{{\omega ^2}}}\frac{2}{E}}  + \mathcal{O}\left( {{E^{1/2}}} \right) .
\end{equation}
The latter means that at the limit $\delta \varphi \to 0$,
\begin{equation}
a \simeq \frac{\kappa }{{\delta \varphi }},
\label{eq:a_small_df}
\end{equation}
where
\begin{equation}
\kappa  = - 16{\Lambda ^4}{\zeta ^2}\left( {{\omega _1};{\Lambda ^{ - 4}},0} \right)\left( {1 + \frac{1}{{{\omega ^2}}}} \right) = - 16{\Lambda ^2}{\zeta ^2}\left( {{\omega _1};1,0} \right)\left( {1 + \frac{1}{{{\omega ^2}}}} \right) .
\label{eq:kappa}
\end{equation}

At the limit of the cusps ($\omega \to \infty$), the parameter $\kappa$ assumes the value
\begin{equation}
\kappa  = - 16{\Lambda ^2}{\zeta ^2}\left( {{\omega _1};1,0} \right) .
\end{equation}

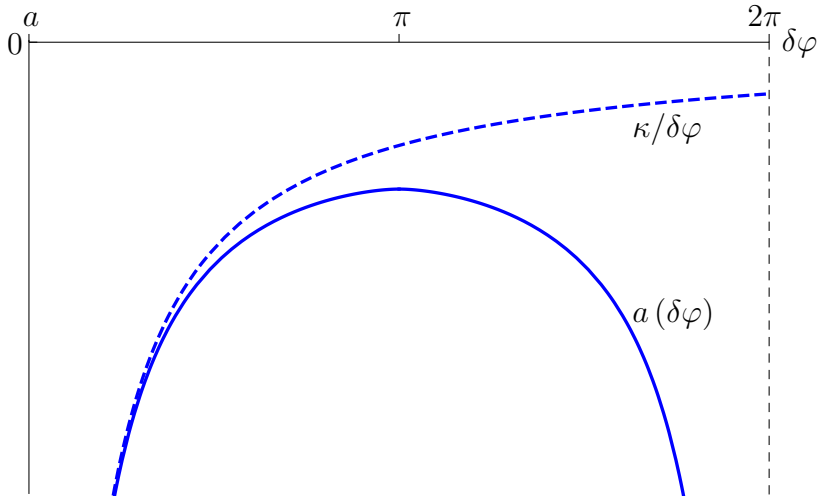
\begin{figure}[ht]
\centering
\begin{picture}(65,42)
\put(1,1){\includegraphics[width = 0.6\textwidth]{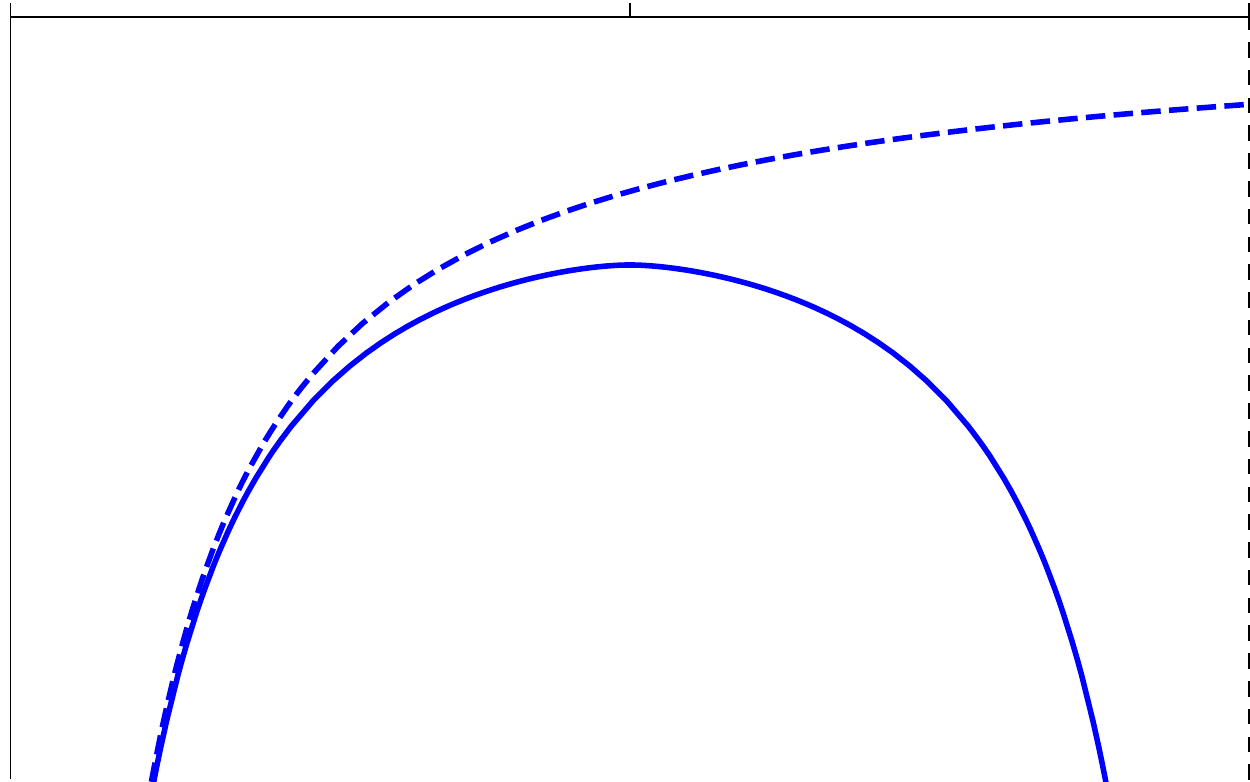}}
\put(1,39){$a$}
\put(-0.25,36.75){$0$}
\put(30.75,39){$\pi$}
\put(59.25,39){$2\pi$}
\put(62,37){$\delta \varphi$}
\put(50,15){$a \left(\delta \varphi \right)$}
\put(50,30){$\kappa / \delta \varphi$}
\end{picture}
\vspace{-5pt}
\caption{The coefficient of the logarithmic term and its asymptotic behaviour for small values of $\delta \varphi$}
\label{fig:small_angle}
\end{figure}

In figure \ref{fig:small_angle}, the dependence of the coefficient $a$ on the parameter $\delta \varphi$ is shown and compared to the asymptotic form for small values of $\delta \varphi$ \eqref{eq:a_small_df}.

\subsection{The Definition of the Angle}
\label{subsec:angle_definition}

The expansion of the holographic entanglement entropy in the cut-off energy scale in the case of elliptic minimal surfaces contains the same terms as in the case of a cusp: a leading ``area law`` term, a constant term and a universal logarithmic term due to the non-smooth points. Should we expect such an expansion in the case of elliptic minimal surfaces? Naively assuming that our case corresponds to a cusp of vanishing angular opening, we should not; we know that the logarithmic term for cusps diverges as $1 / \Omega$ for small angles, and, thus, we should expect divergences of a higher order like $L \ln L$ in the case of elliptic minimal surfaces. On the contrary, we have shown that such a more divergent term does not appear and furthermore the logarithmic term diverges as $\kappa / \delta \varphi$ for small $\delta \varphi$ similarly to the case of cusps for small angular openings, with the role of the angle $\Omega$ being played by the parameter $\delta \varphi$. This implies that the geometrical feature affecting the logarithmic term is not exactly the angle; The appropriate geometrical feature should be carefully defined so that its definition is equivalent to the definition of the angle in the case of cusps, but it differs in the case of spiral spikes, coinciding with the value of the parameter $\delta \varphi$.

The common definition of the angular opening requires the use of a circle centred at the singular point. Then, the naive definition of the angle is the ratio of the corresponding arc length divided by the radius of the circle in the limit the radius of the circle goes to zero
\begin{equation}
\Omega : = \mathop {\lim }\limits_{{r_0} \to 0} \frac{{{\ell _{\rm{arc}}}}}{{{r_0}}} .
\label{eq:angle_def_1}
\end{equation}
However, one could have defined the angle based on the area of the corresponding sector as,
\begin{equation}
\tilde \Omega : = 2\pi \mathop {\lim }\limits_{{r_0} \to 0} \frac{{{A_{\rm{sector}}}}}{{\pi r_0^2}} .
\label{eq:angle_def_2}
\end{equation}
\begin{figure}[ht]
\centering
\begin{picture}(45,45)
\put(0,0){\includegraphics[width = 0.45\textwidth]{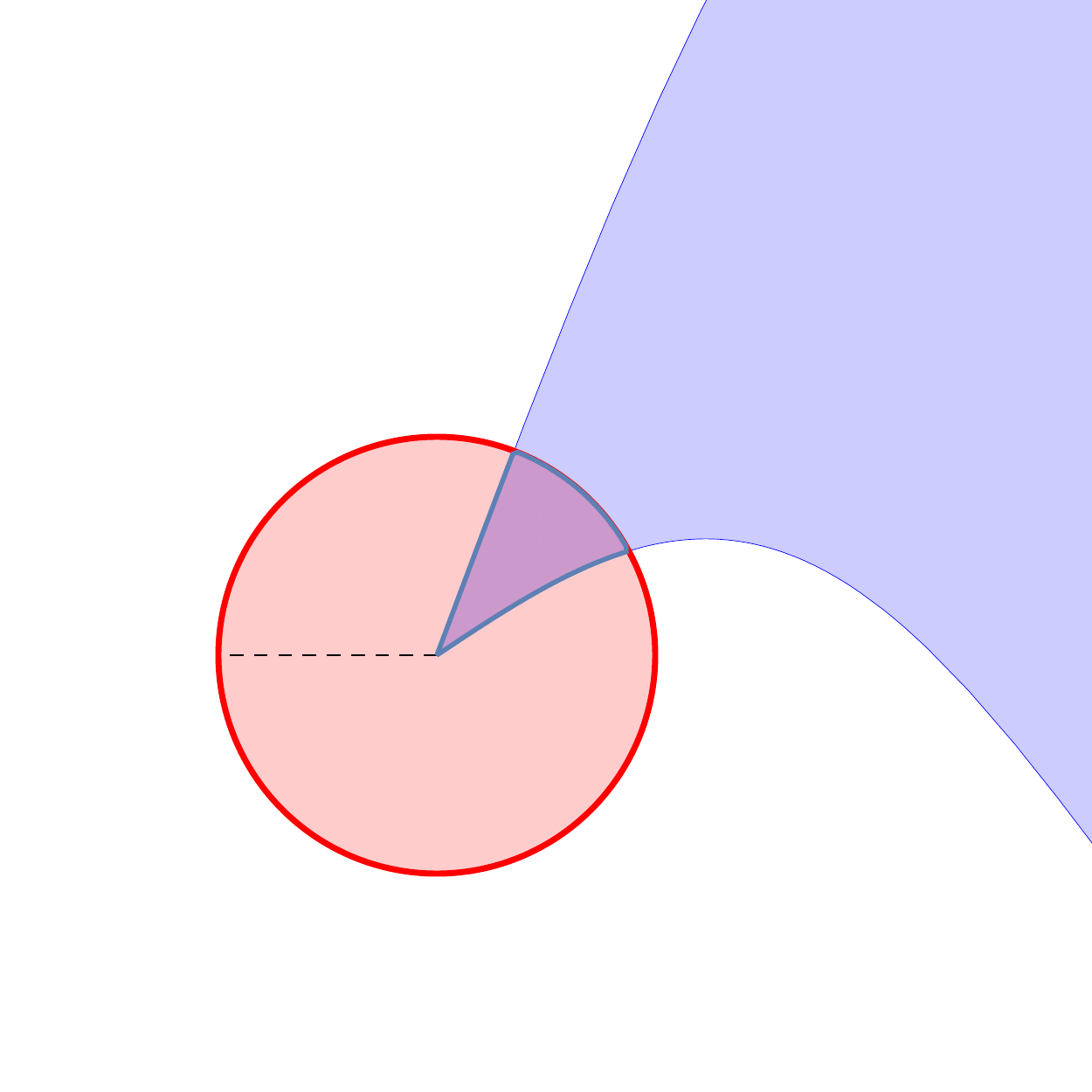}}
\put(12.5,19){$r_0$}
\put(23.75,26){${\ell _{\rm{arc}}}$}
\put(18.75,22){${A_{\rm{sector}}}$}
\end{picture}
\vspace{-40pt}
\caption{The two possible definitions of the angle $\Omega$}
\label{fig:angle_def_basic}
\end{figure}
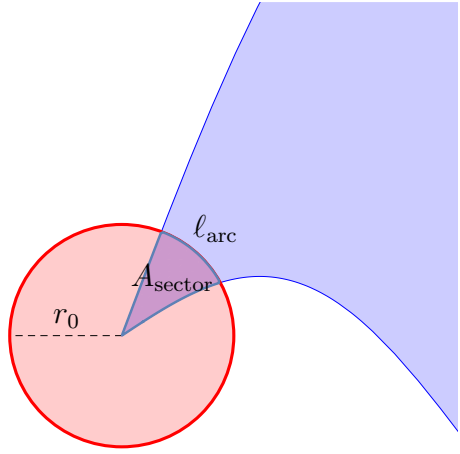

\begin{figure}[ht]
\centering
\begin{picture}(80,73)
\put(1,41){\includegraphics[width = 0.32\textwidth]{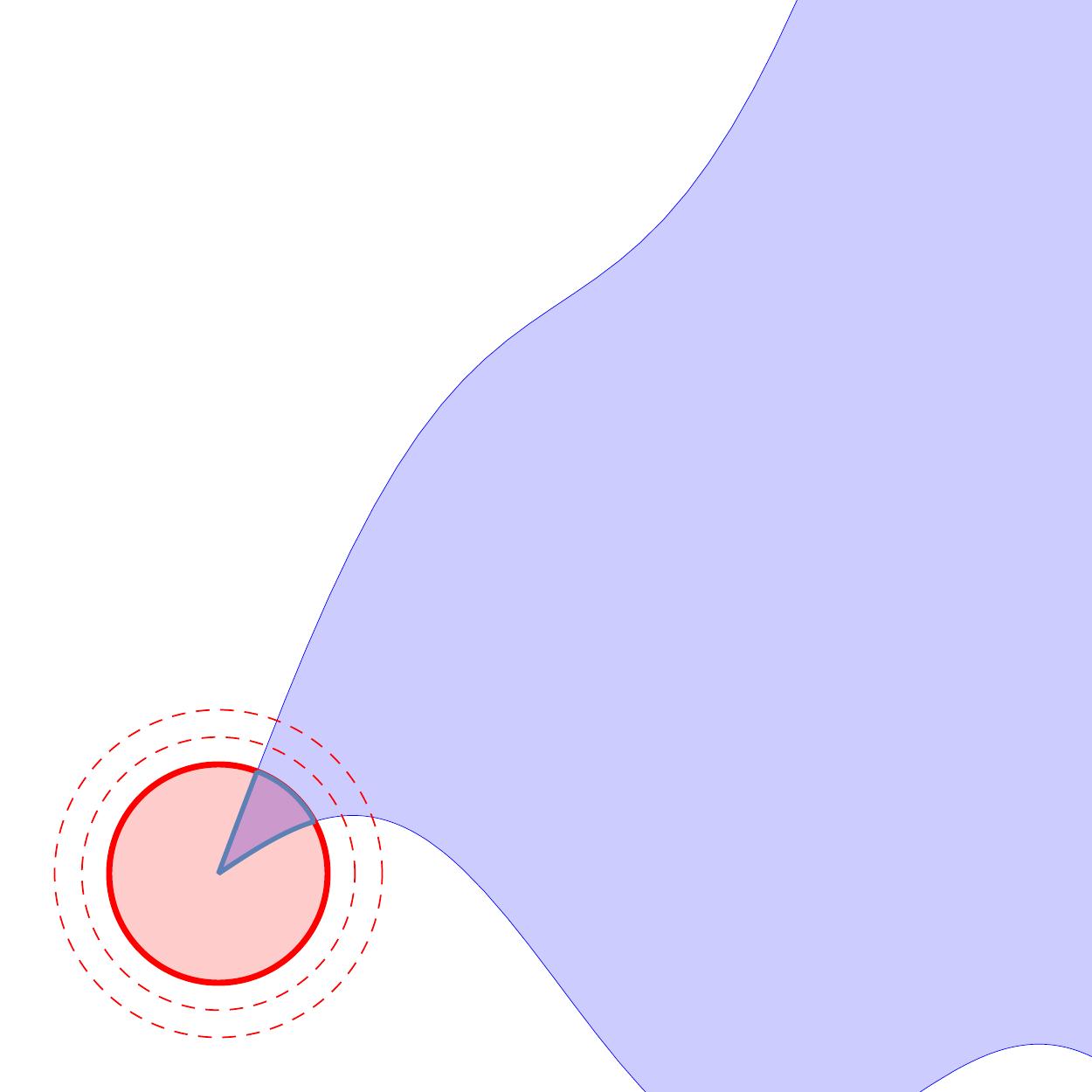}}
\put(44,41){\includegraphics[width = 0.32\textwidth]{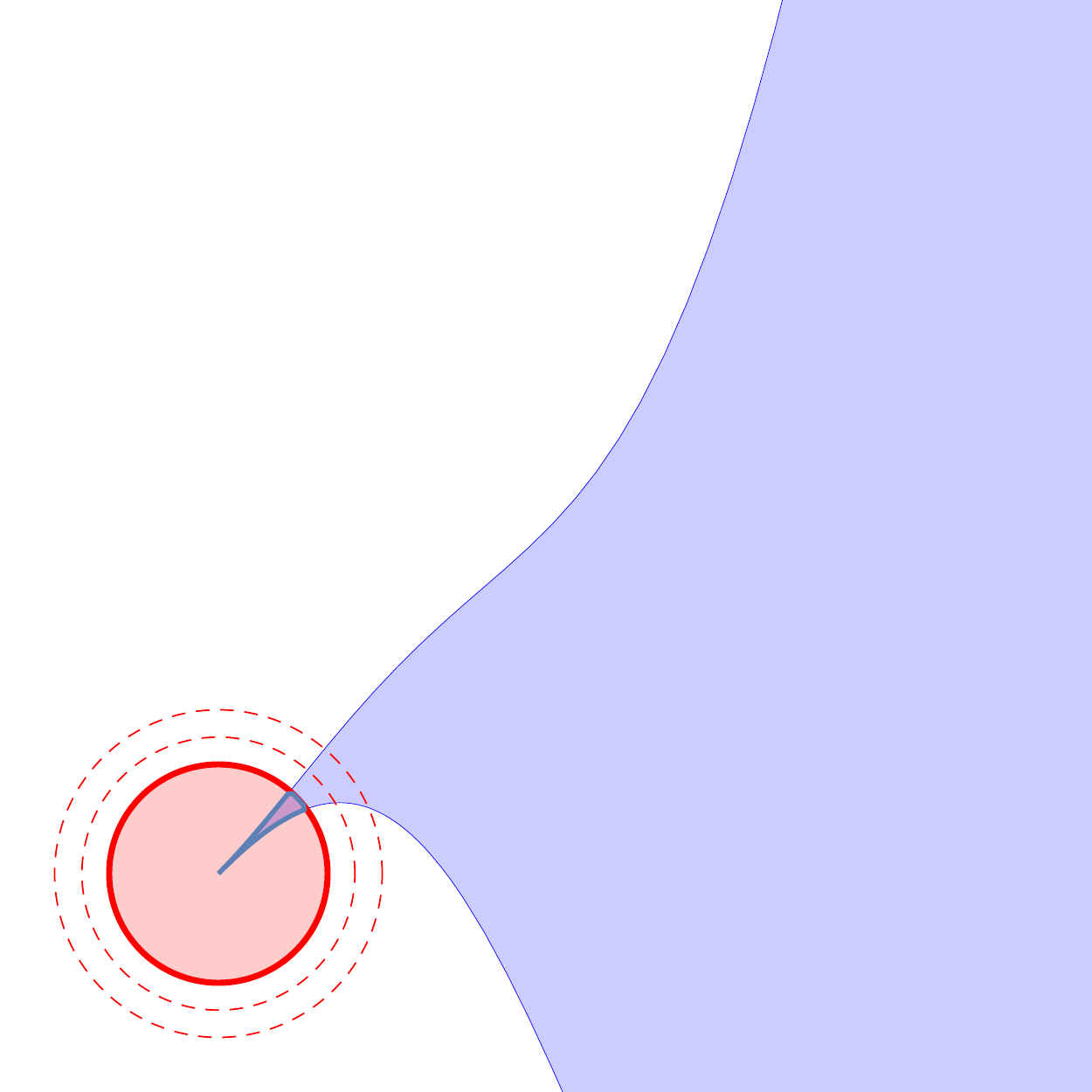}}
\put(22.5,3){\includegraphics[width = 0.32\textwidth]{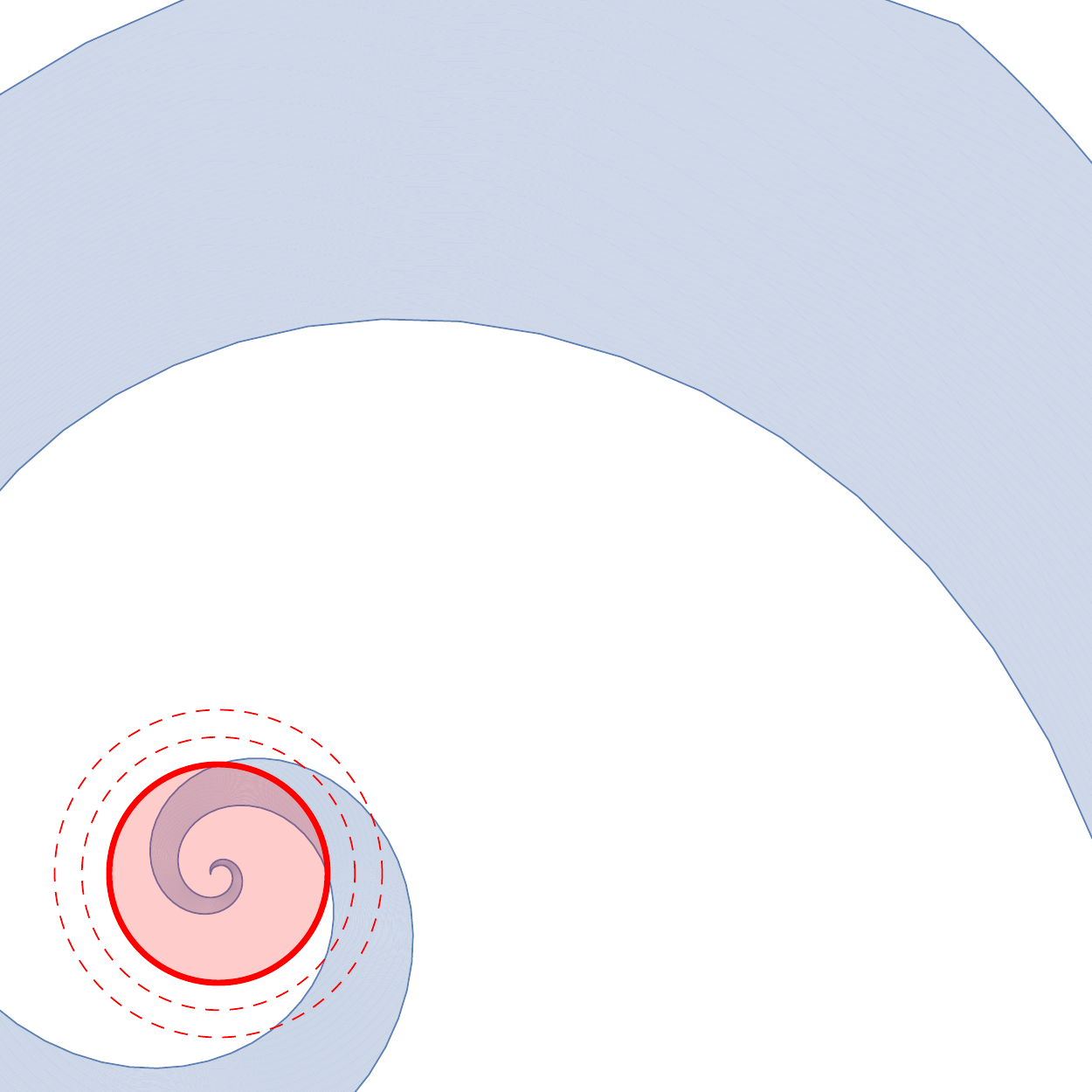}}
\put(15,38){cusp}
\put(58,38){spike}
\put(32.5,0){spiral spike}
\end{picture}
\caption{The area definition of the angle in the case of a cusp, a spike and the case of a spiral spike that appears in the entangling curve of elliptic minimal surfaces}
\label{fig:angle_def}
\end{figure}
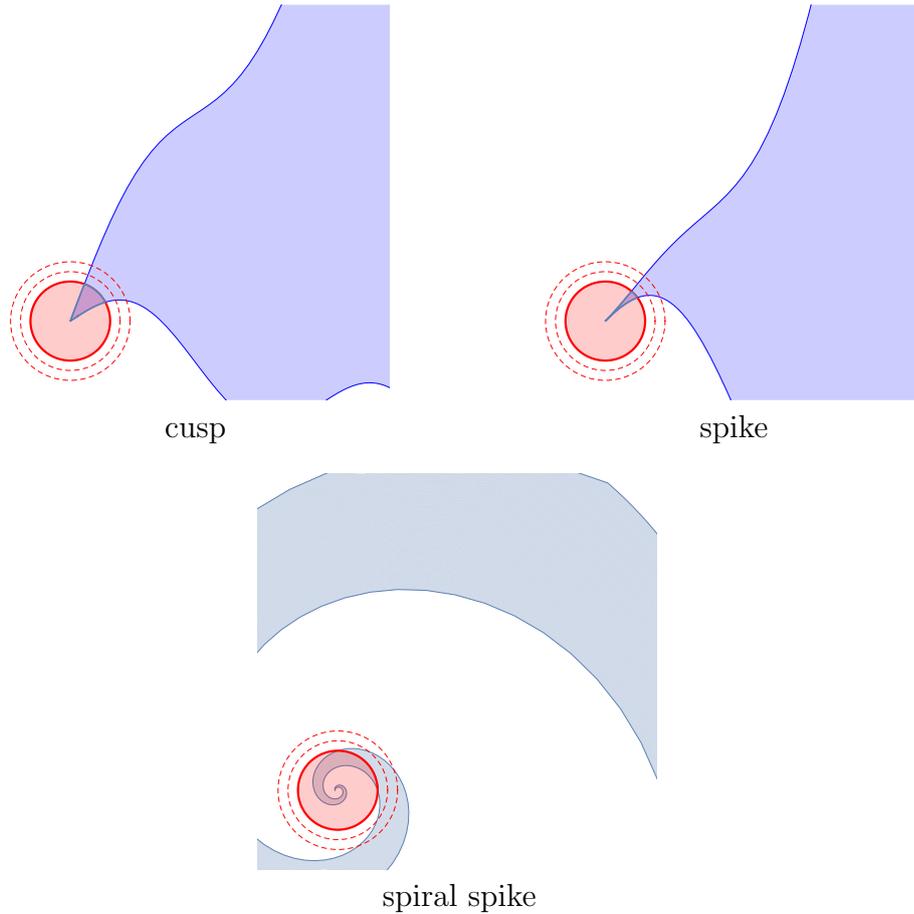

Of course in most cases the two definitions \eqref{eq:angle_def_1} and \eqref{eq:angle_def_2} give identical results. However, this is not the case for elliptic minimal surfaces and their spiral spikes. This strange behaviour is permitted by the self-similarity of the entangling curve and by the fact that the length of the segment of the entangling curve from the singular spiky point to any other point is infinite. In figure \ref{fig:angle_def}, three cases of singular points are depicted: a cusp (top left), a spike (top right) and a spiral spike of the entangling curves under consideration (bottom).

Trivially, in the case of a cusp,
\begin{equation}
{{2\pi }} \mathop {\lim }\limits_{r_0 \to 0 } \frac{{A_{\rm{sector}}}}{{\pi r_0^2}} = {\Omega } = \mathop {\lim }\limits_{{r_0} \to 0} \frac{{{\ell _{\rm{arc}}}}}{{{r_0}}}.
\end{equation}
In the limit that the cusp angular opening goes to zero and the cusp degenerates to a spike, we get
\begin{equation}
{{2\pi }} \mathop {\lim }\limits_{r_0 \to 0 } \frac{{A_{\rm{sector}}}}{{\pi r_0^2}} = 0 = \mathop {\lim }\limits_{{r_0} \to 0} \frac{{{\ell _{\rm{arc}}}}}{{{r_0}}}.
\end{equation}
However, in the singular points of the elliptic entangling curves, although the angular opening is not well-defined,
\begin{equation}
{{2\pi }} \mathop {\lim }\limits_{r_0 \to 0 } \frac{{A_{\rm{sector}}}}{{\pi r_0^2}} = {\delta \varphi } \neq \mathop {\lim }\limits_{{r_0} \to 0} \frac{{{\ell _{\rm{arc}}}}}{{{r_0}}} ,
\end{equation}
and, thus, the two definitions do not lead to identical results.

As we discussed in the beginning of this section, the appropriate definition for the geometric feature that determines the coefficients of the universal terms arising from the existence of singular non-smooth points of the entangling curve, should coincide with the angle in the case of cusps and with the parameter $\delta \varphi$ in the case of spiral spikes. It follows that the appropriate definition is not the arc length definition \eqref{eq:angle_def_1}, but the area definition \eqref{eq:angle_def_2}.

The definition \eqref{eq:angle_def_2} is also more natural as seen from the perspective of the physics of entanglement. Entanglement entropy directly depends on the way the entangling curve separates the degrees of freedom. The number of the latter is proportional to the area of the sector and not the length of the arc. In this language, a smooth entangling curve has the special property of dividing the degrees of freedom in the neighbourhood of any of its points to two equal subsets.

\subsection{Smoothness of the Logarithmic Term Coefficient}
\label{subsec:instabilities}

In the case of cusps, the demand that the coefficient of the logarithmic term $a$ is a smooth function of the angular opening $\Omega$, combined with the symmetry property \eqref{eq:a_log_symmetry} implies that the Taylor expansion of $a$ around $\Omega = \pi$ contains no first order term. Actually, the zeroth order term also vanishes, since at $\Omega = \pi$ the entangling curve is smooth and no logarithmic term occurs. In the case of elliptic minimal surfaces, the latter is obviously not true, but should we expect that the former still holds at $\delta \varphi = \pi$?

There is one more demand that should have been posted to ensure that $\Omega = \pi$ is a smooth extremum of the coefficient of the logarithmic term: the demand that there is a unique minimal surface for the given entangling curve. When this is not the case, the dependence of $a$ on $\delta \varphi$ is not one-to-one. Although we would expect that plotting $a$ as function of $\delta \varphi$ would be again a smooth curve, the latter will have self-intersections. Therefore, the segment of the above curve that corresponds to the globally stable minimal surfaces, is not guaranteed to be smooth, but it may present singular points, being self-intersections of the smooth curve.

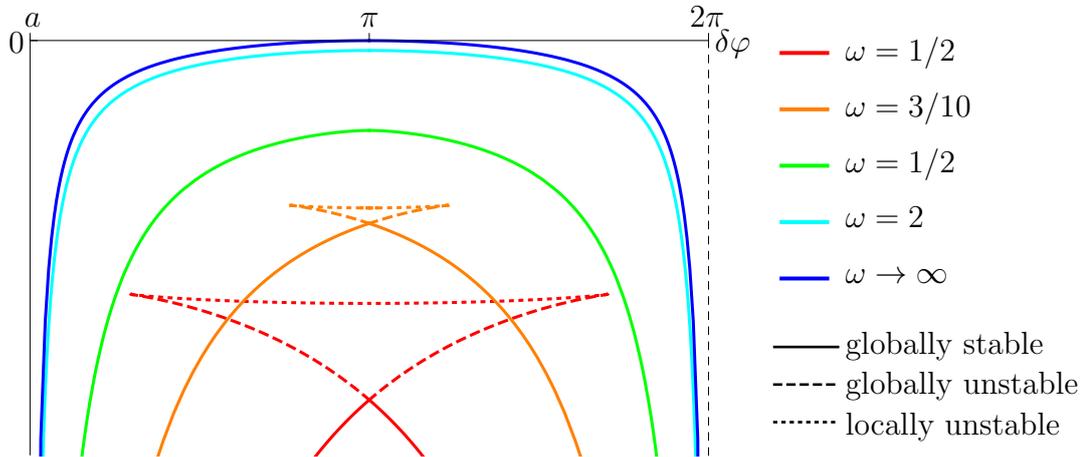
\begin{figure}[hbt]
\centering
\begin{picture}(85,36.25)
\put(1,0){\includegraphics[width = 0.55\textwidth]{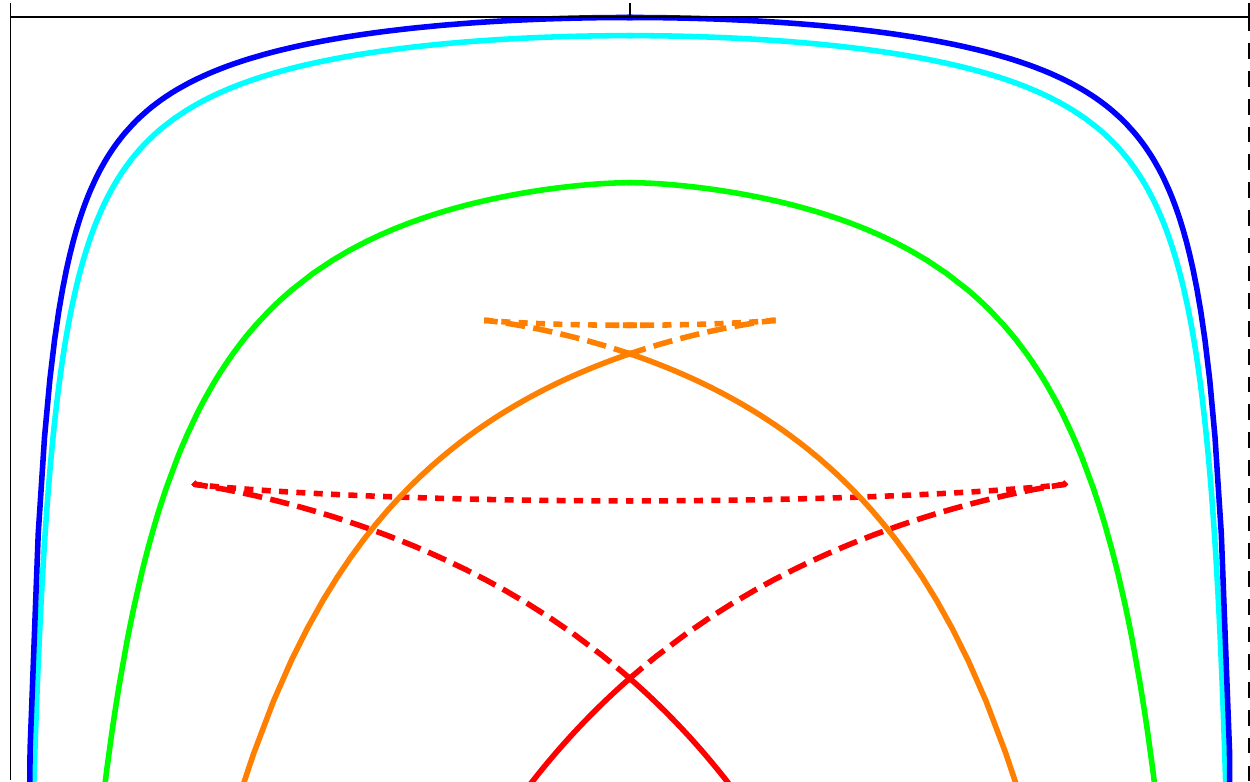}}
\put(60.625,11.25){\includegraphics[width = 0.0775\textwidth]{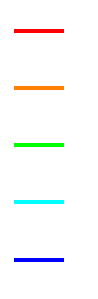}}
\put(60.5,0.75){\includegraphics[width = 0.0775\textwidth]{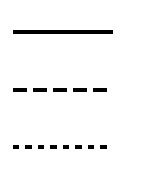}}
\put(1,34.75){$a$}
\put(-0.25,32.625){$0$}
\put(28,34.75){$\pi$}
\put(54.5,34.75){$2\pi$}
\put(56.5,32.75){$\delta \varphi$}
\put(67,14){$\omega \to \infty$}
\put(67,18.5){$\omega = 2$}
\put(67,23){$\omega = 1/2$}
\put(67,27.5){$\omega = 3/10$}
\put(67,32){$\omega = 1/2$}
\put(67,1.75){locally unstable}
\put(67,5){globally unstable}
\put(67,8.25){globally stable}
\end{picture}
\vspace{-3pt}
\caption{The coefficient of the logarithmic term, as function of the parameter $\delta \varphi$, for various values of $\omega$}
\label{fig:angle_pi}
\end{figure}

The demand of the existence of a unique minimal surface is satisfied in the case of cusps. However, it is not always true for more general elliptic minimal surfaces. As shown in \cite{Pastras:2016vqu}, there is a critical value of the parameter $\omega$, equal to $\omega_{\rm{cr}} \simeq 0.458787$, so that for $\omega <  \omega_{\rm{cr}}$, there are in general more than one minimal surfaces for the same entangling curve. Indeed, as shown in figure \ref{fig:angle_pi}, if only globally stable configurations are considered, when $\omega <  \omega_{\rm{cr}}$ the coefficient of the logarithmic term has an extremum at $\delta \varphi = \pi$, but not a smooth one. The smooth extremum in such cases corresponds to the helicoid limit of the minimal surfaces, which is globally \cite{Pastras:2016vqu} and locally \cite{Wang_helicoids} unstable for these values of $\omega$.

\setcounter{equation}{0}
\section{Discussion}
\label{sec:discussion}

The elliptic minimal surfaces in AdS$_4$ correspond to entangling curves that contain singular non-smooth points that generalize cusps. For these surfaces, the holographic entanglement entropy contains a logarithmic term that depends solely on the geometry of the singular points, in a similar manner to the corresponding terms in the case of cusps. This logarithmic term shares some of the properties of the corresponding terms in the case of cusps, but not all of them.

The entangling curves corresponding to elliptic minimal surfaces contain two singular non-smooth points. Although an angle cannot be well defined in these points, it is clear that the entangling curve divides the degrees of freedom in its neighbourhood to two unequal subsets, similarly to the case of a cusps. The number of the degrees of freedom in each of the two subsets is determined by the elliptic minimal surface  parameter $\delta \varphi$, in the same way it is determined by the angle $\Omega$ in the case of cusps.

The similarity of the divergence of the coefficient of the logarithmic term for small parameters $\delta \varphi$ to the divergence of the logarithmic term for small angles $\Omega$ in the case of cusps, suggests that the logarithmic term indeed depends only on the geometry of the singular non-smooth point. However, the geometric feature determining these logarithmic terms should not be defined as the angular opening, the ratio of the arc length to the radius, but rather as the ratio of the area sector to the circle area. This is a more natural definition when considering the physics of entanglement and entanglement entropy, as it is directly connected to the separation of the degrees of freedom by the entangling curve. It is more natural to say that the geometric characteristic determining the universal terms due to singular points is not really an angle, ranging between $0$ and $2 \pi$, but rather a ratio describing the separation of the degrees of freedom at the neighbourhood of the singular point, 
\begin{equation}
\lambda := \mathop {\lim }\limits_{r_0 \to 0 } \frac{{{A_{\mathrm{sector} \; \mathrm{within} \; \mathrm{region}\;A}}}}{{\pi r_0^2}} ,
\end{equation}
ranging between $0$ and $1$.

A smooth entangling surface is characterized by $\lambda = 1 / 2$ at all points. In this context, it would be interesting to investigate whether the dependence of entanglement entropy on geometric characteristics of a smooth entangling surface, such as the curvature \cite{Solodukhin:2008dh}, can be described in a similar, unifying manner, considering these characteristics as those determining the way that $\lambda$ tends to $1/2$ at the limit $r_0 \to 0$.

Finally, the logarithmic term does not inherit the property of vanishing at $\delta \varphi = \pi$ from the cusps case. More interestingly, it does not inherit the property of being stationary or even smooth at $\delta \varphi = \pi$. This is a direct consequence of the fact that the minimal surface is not always uniquely defined by the entangling curve, but more than one minimal surfaces corresponding to the same entangling curve may exist.


\subsection*{Acknowledgements}
The research of G.P. is funded by the ``Post-doctoral researchers support'' action of the operational programme ``human resources development, education and long life learning, 2014-2020'', with priority axes 6, 8 and 9, implemented by the Greek State Scholarship Foundation and co-funded by the European Social Fund - ESF and National Resources of Greece.

G.P. would like to thank M. Axenides, E. Floratos and D. Katsinis for useful discussions.

\appendix

\renewcommand{\thesection}{\Alph{section}}
\renewcommand{\thesubsection}{\Alph{section}.\arabic{subsection}}
\renewcommand{\theequation}{\Alph{section}.\arabic{equation}}

\setcounter{equation}{0}
\section{Formulae on Elliptic Minimal Surfaces}
\label{sec:formulae}

In this appendix we review some features of the elliptic minimal surfaces used in the calculations throughout the main text. More details on the construction of the static elliptic minimal surfaces, their properties and their relation with the Pohlmeyer reduced integrable system are provided in \cite{Pastras:2016vqu}.

In global coordinates, where the bulk metric assumes the form
\begin{equation}
d{s^2} = - {\left( {1 + \frac{{{r^2}}}{{{\Lambda ^2}}}} \right)}d{t^2} + {\left( {1 + \frac{{{r^2}}}{{{\Lambda ^2}}}} \right)^{ - 1}}d{r^2} + {r^2}\left( {d{\theta ^2} + {{\sin }^2}\theta d{\varphi ^2}} \right) ,
\end{equation}
the elliptic minimal surfaces can be written in the following parametric form in terms of the isothermal coordinates $u$ and $v$,
\begin{align}
r &= \Lambda \sqrt {\frac{{\wp \left( u \right) - \wp \left( {{a_1}} \right)}}{{\wp \left( {{a_2}} \right) - \wp \left( {{a_1}} \right)}}{{\cosh }^2}\left( {{\ell _1}v + {\varphi _1}\left( u \right)} \right) - 1} , \label{eq:app_r}\\
\theta  &= {\tan ^{ - 1}}\left( {\sqrt {\frac{{\wp \left( u \right) - \wp \left( {{a_1}} \right)}}{{\wp \left( u \right) - \wp \left( {{a_2}} \right)}}} \csch\left( {{\ell _1}v + {\varphi _1}\left( u \right)} \right)} \right) , \label{eq:app_t}\\
\varphi  &= {\ell _2}v - {\varphi _2}\left( u \right) . \label{eq:app_p}
\end{align}
These surfaces are a two-parameter familly of minimal surfaces. They are naturally parametrized in terms of the quantity $E$, which specifies the corresponding solution of the Pohlmeyer reduced problem, and $\wp \left( {{a_1}} \right)$.

The moduli of the Weierstrass elliptic and related functions appearing in the above expressions are equal to
\begin{align}
{g_2} &= \frac{{{E^2}}}{3} + \frac{1}{{{\Lambda ^4}}} , \label{eq:Elliptic_g2}\\
{g_3} &=  - \frac{E}{3}\left( {\frac{{{E^2}}}{9} + \frac{1}{{2{\Lambda ^4}}}} \right) . \label{eq:Elliptic_g3}
\end{align}
The above values of the moduli imply that the associated cubic polynomial has three real roots independently of the value of the constant $E$. These roots equal
\begin{align}
{e_1} &=  - \frac{E}{{12}} + \frac{1}{4}\sqrt {{E^2} + \frac{4}{{{\Lambda ^4}}}} , \label{eq:Elliptic_e1}\\
{e_2} &= \frac{E}{6} , \label{eq:Elliptic_e2}\\
{e_3} &=  - \frac{E}{{12}} - \frac{1}{4}\sqrt {{E^2} + \frac{4}{{{\Lambda ^4}}}} .\label{eq:Elliptic_e3}
\end{align}
The functions ${\varphi _1}$ and ${\varphi _2}$ appearing in the equations \eqref{eq:app_r}, \eqref{eq:app_t} and \eqref{eq:app_p} are given by
\begin{align}
{\varphi _1}\left( u \right) &= \frac{1}{2}\ln \left( { - \frac{{\sigma \left( {u + {a_1}} \right)}}{{\sigma \left( {u - {a_1}} \right)}}} \right) - \zeta \left( {{a_1}} \right)u ,\\
{\varphi _2}\left( u \right) &=  - \frac{i}{2}\ln \left( { - \frac{{\sigma \left( {u + {a_2}} \right)}}{{\sigma \left( {u - {a_2}} \right)}}} \right) + i\zeta \left( {{a_2}} \right)u ,
\end{align}
where $\zeta$ and $\sigma$ are the Weierstrass zeta and sigma functions respectively. Finally, the parameters $\ell _1$ and $\ell _2$ appearing in the equations \eqref{eq:app_r}, \eqref{eq:app_t} and \eqref{eq:app_p} equal
\begin{align}
\ell _1^2 &=  - \wp \left( {{a_1}} \right) - 2{e_2} , \label{eq:construction_cosh_a1}\\
\ell _2^2 &=  \wp \left( {{a_2}} \right) + 2{e_2} , \label{eq:construction_cosh_a2}
\end{align}
where the parameters $a_1$ and $a_2$ must be chosen so that they satisfy
\begin{equation}
\wp \left( {{a_1}} \right) + \wp \left( {{a_2}} \right) =  - {e_2} .
\label{eq:construction_cosh_Virasoro_final}
\end{equation}

The entangling curve corresponding to the minimal surface described by the equations \eqref{eq:app_r}, \eqref{eq:app_t} and \eqref{eq:app_p} is completely determined by the parameters $\omega$ and $\delta \varphi$, which are functions of $E$ and $\wp \left( {{a_1}} \right)$. The entangling curve can be specified analytically taking the limits $u \to 2 n \omega_1$ and $u \to 2 \left( n + 1 \right) \omega_1$. It turns out that it is the union of two spiral curves of the form
\begin{align}
\cot {\theta _ + } &= \sinh \left( {\omega\left( {{\varphi _ + } + {\varphi _0}} \right)} \right) ,\\
\cot {\theta _ - } &= \sinh \left( {\omega\left( {{\varphi _ - } + {\varphi _0} - \delta \varphi } \right)} \right) .
\end{align}
The parameters $\omega$ and $\delta \varphi$ are equal to
\begin{align}
\omega &= \frac{{{\ell _1}}}{{{\ell _2}}}, \label{eq:boundary_omega}\\
\delta \varphi &= \pi  - 2\left( {{\mathop{\rm Im}\nolimits} {\delta _2} + \frac{{{\ell _2}}}{{{\ell _1}}}{\mathop{\rm Re}\nolimits} {\delta _1}} \right) , \label{eq:boundary_df}
\end{align}
where $\delta_1$ and $\delta_2$ are given by
\begin{align}
{\delta _1} &\equiv \zeta \left( {{\omega _1}} \right){a_1} - \zeta \left( {{a_1}} \right){\omega _1} ,\\
{\delta _2} &\equiv \zeta \left( {{\omega _1}} \right){a_2} - \zeta \left( {{a_2}} \right){\omega _1} .
\end{align}

.


\begin{thebibliography}{1}


\bibitem{Ryu:2006bv}
  S.~Ryu and T.~Takayanagi,
  ``Holographic Derivation of Entanglement Entropy from AdS/CFT",
  Phys.\ Rev.\ Lett.\  {\bf 96}, 181602 (2006)
  [hep-th/0603001].
  
\bibitem{Ryu:2006ef}
  S.~Ryu and T.~Takayanagi,
  ``Aspects of Holographic Entanglement Entropy",
  JHEP {\bf 0608}, 045 (2006)
  [hep-th/0605073].


\bibitem{Hubeny:2007xt}
  V.~E.~Hubeny, M.~Rangamani and T.~Takayanagi,
  ``A Covariant Holographic Entanglement Entropy Proposal",
  JHEP {\bf 0707}, 062 (2007)
  [arXiv:0705.0016 [hep-th]].
\bibitem{Nishioka:2009un}
  T.~Nishioka, S.~Ryu and T.~Takayanagi,
  ``Holographic Entanglement Entropy: An Overview",
  J.\ Phys.\ A {\bf 42}, 504008 (2009)
  [arXiv:0905.0932 [hep-th]].
\bibitem{VanRaamsdonk:2009ar}
  M.~Van Raamsdonk,
  ``Comments on Quantum Gravity and Entanglement",
  arXiv:0907.2939 [hep-th].
\bibitem{VanRaamsdonk:2010pw}
  M.~Van Raamsdonk,
 ``Building up Spacetime with Quantum Entanglement",
  Gen.\ Rel.\ Grav.\  {\bf 42}, 2323 (2010)
  [Int.\ J.\ Mod.\ Phys.\ D {\bf 19}, 2429 (2010)]
  [arXiv:1005.3035 [hep-th]].
\bibitem{Takayanagi:2012kg}
  T.~Takayanagi,
  ``Entanglement Entropy from a Holographic Viewpoint",
  Class.\ Quant.\ Grav.\  {\bf 29}, 153001 (2012)
  [arXiv:1204.2450 [gr-qc]].

\bibitem{Blanco:2013joa}
  D.~D.~Blanco, H.~Casini, L.~Y.~Hung and R.~C.~Myers,
  ``Relative Entropy and Holography",
  JHEP {\bf 1308}, 060 (2013)
  [arXiv:1305.3182 [hep-th]].

\bibitem{Wong:2013gua}
  G.~Wong, I.~Klich, L.~A.~Pando Zayas and D.~Vaman,
  ``Entanglement Temperature and Entanglement Entropy of Excited States",
  JHEP {\bf 1312}, 020 (2013)
  [arXiv:1305.3291 [hep-th]].


\bibitem{Bombelli:1986rw} 
  L.~Bombelli, R.~K.~Koul, J.~Lee and R.~D.~Sorkin,
  ``A Quantum Source of Entropy for Black Holes'',
  Phys.\ Rev.\ D {\bf 34}, 373 (1986).
  doi:10.1103/PhysRevD.34.373

\bibitem{Srednicki:1993im} 
  M.~Srednicki,
  ``Entropy and Area'',
  Phys.\ Rev.\ Lett.\  {\bf 71}, 666 (1993)
  doi:10.1103/PhysRevLett.71.666
  [hep-th/9303048].
  
  
  \bibitem{Casini:2011kv}
  H.~Casini, M.~Huerta and R.~C.~Myers,
  ``Towards a Derivation of Holographic Entanglement Entropy",
  JHEP {\bf 1105}, 036 (2011)
  [arXiv:1102.0440 [hep-th]].  
    
  \bibitem{Myers:2010xs}
  R.~C.~Myers and A.~Sinha,
  ``Seeing a c-theorem with Holography",
  Phys.\ Rev.\ D {\bf 82}, 046006 (2010)
  [arXiv:1006.1263 [hep-th]].

\bibitem{Myers:2010tj}
  R.~C.~Myers and A.~Sinha,
  ``Holographic c-theorems in Arbitrary Dimensions",
  JHEP {\bf 1101}, 125 (2011)
  [arXiv:1011.5819 [hep-th]].

\bibitem{Solodukhin:2008dh}
  S.~N.~Solodukhin,
  ``Entanglement Entropy, Conformal Invariance and Extrinsic Geometry",
  Phys.\ Lett.\ B {\bf 665}, 305 (2008)
  [arXiv:0802.3117 [hep-th]].


\bibitem{Bueno:2015xda} 
  P.~Bueno and R.~C.~Myers,
  ``Corner Contributions to Holographic Entanglement Entropy'',
  JHEP {\bf 1508}, 068 (2015)
  doi:10.1007/JHEP08(2015)068
  [arXiv:1505.07842 [hep-th]].
  
\bibitem{Bueno:2015rda} 
  P.~Bueno, R.~C.~Myers and W.~Witczak-Krempa,
  ``Universality of Corner Entanglement in Conformal Field Theories'',
  Phys.\ Rev.\ Lett.\  {\bf 115}, 021602 (2015)
  doi:10.1103/PhysRevLett.115.021602
  [arXiv:1505.04804 [hep-th]].

\bibitem{Bueno:2015qya} 
  P.~Bueno, R.~C.~Myers and W.~Witczak-Krempa,
  ``Universal Corner Entanglement from Twist Operators'',
  JHEP {\bf 1509}, 091 (2015)
  doi:10.1007/JHEP09(2015)091
  [arXiv:1507.06997 [hep-th]].

\bibitem{Bueno:2015lza} 
  P.~Bueno and R.~C.~Myers,
  ``Universal Entanglement for Higher Dimensional Cones'',
  JHEP {\bf 1512}, 168 (2015)
  doi:10.1007/JHEP12(2015)168
  [arXiv:1508.00587 [hep-th]].
  

\bibitem{Pastras:2016vqu} 
  G.~Pastras,
  ``Static Elliptic Minimal Surfaces in AdS(4)'',
  arXiv:1612.03631 [hep-th].
  
\bibitem{Pastras:2017afl} 
  G.~Pastras,
  ``Elliptic String Solutions in AdS(3) and Elliptic Minimal Surfaces in AdS(4),''
  arXiv:1710.00545 [hep-th].

\bibitem{Wang_helicoids}
  B.~Wang, 
  ``Stability of Helicoids in Hyperbolic Three-Dimensional Space'',
  [arXiv:1502.04764[math.DG]].


\end{thebibliography}
\end{document}